\documentclass[prd,onecolumn,superscriptaddress,showpacs,nofootinbib,preprintnumbers]{revtex4}
\usepackage{amsmath}
\usepackage{amsfonts}
\usepackage{graphicx}
\usepackage{dcolumn}
\usepackage{hyperref}
\usepackage{bm}
\def\rd{{\rm d}}

\def\be{\begin{equation}}
\def\ee{\end{equation}}
\def\bea{\begin{eqnarray}}
\def\eea{\end{eqnarray}}

\def\5{\overline 5}

\begin{document}

\title{Second-order matter density perturbations and skewness \\
in scalar-tensor modified gravity models}

\author{Takayuki Tatekawa}
\affiliation{Department of Computer Science, Kogakuin University,
1-24-2 Nishi-shinjuku, Shinjuku, Tokyo, 163-8677 Japan}
\email{tatekawa@cpd.kogakuin.ac.jp}
\affiliation{Research Institute for Science and Engineering,
Waseda University,
3-4-1 Okubo, Shinjuku, Tokyo, 169-8555 Japan}
\affiliation{Department of Physics, Ochanomizu University,
2-1-1 Otsuka, Bunkyo, Tokyo, 112-8610, Japan}

\author{Shinji Tsujikawa}
\affiliation{Department of Physics, Faculty of Science, Tokyo University of Science, 
1-3, Kagurazaka, Shinjuku-ku, Tokyo 162-8601, Japan}
\email{shinji@rs.kagu.tus.ac.jp}

\date{\today}

\vskip 1pc

%-------------------
\begin{abstract}

We study second-order cosmological perturbations in scalar-tensor
models of dark energy that satisfy local gravity constraints, including 
$f(R)$ gravity. We derive equations for matter fluctuations
under a sub-horizon approximation and clarify conditions 
under which first-order perturbations in the scalar field 
can be neglected relative to second-order matter and velocity 
perturbations.
We also compute the skewness of the matter density distribution
and find that the difference from the $\Lambda$CDM model is 
only less than a few percent even if the growth rate of first-order perturbations 
is significantly different from that in the $\Lambda$CDM model.
This shows that the skewness provides a model-independent test
for the picture of gravitational instability from 
Gaussian initial perturbations including scalar-tensor 
modified gravity models.

\end{abstract}
%-------------------


\maketitle
\vskip 1pc

%%%%%%%%%%%%%%%%%%%%%%%%%%
\section{Introduction}
%%%%%%%%%%%%%%%%%%%%%%%%%%

The constantly accumulating observational data \cite{Perl} continue to 
confirm that the Universe has entered the phase of an accelerated
expansion after the matter-dominated epoch.
To reveal the origin of dark energy (DE) responsible for 
this late-time acceleration is one of the most serious stumbling block 
in modern cosmology \cite{review,CST}. 
The first step toward understanding the nature of DE
is to find a signature whether it originates from some modification 
of gravity or it comes from some exotic matter with negative pressure.
If gravity is modified from Einstein's General Relativity, this leaves 
a number of interesting experimental and observational signatures
that can be tested. Especially local gravity experiments generally place 
tight bounds for the parameter space of modified gravity models.

So far many modified gravity DE models have been proposed--
ranging from $f(R)$ gravity \cite{fR} ($R$ is a Ricci scalar), 
scalar-tensor theory \cite{stensor,Boi} to braneworld 
scenarios \cite{brane}. 
The $f(R)$ gravity is presumably the simplest generalization to  
the $\Lambda$-Cold Dark Matter ($\Lambda$CDM) model
($f(R)=R-\Lambda$).
Nevertheless it is generally not easy to construct viable $f(R)$
models that satisfy all stability, experimental and observational constraints
while at the same time showing appreciable deviations from the 
$\Lambda$CDM model. 
In order to avoid that a scalar degree of freedom (scalaron) 
as well as a graviton becomes ghosts or tachyons we require the conditions 
$f_{,RR}>0$ and $f_{,R}>0$ \cite{Star07}.
These conditions are also needed for the stability of density 
perturbations \cite{staper}.
For the existence of a matter-dominated epoch followed by a late-time
acceleration, the models need to be close to the $\Lambda$CDM model
($m \equiv Rf_{,RR}/f_{,R} \approx +0$) in the region $R \gg R_0$
($R_0$ is the present cosmological Ricci scalar) \cite{AGPT}.
Moreover the mass of the scalaron field in the region $R \gg R_0$
is sufficiently heavy for the compatibility with local gravity 
experiments \cite{Nava,CT,TUT,Brax}.
Finally, for the presence of a stable de Sitter fixed point at 
$r \equiv -Rf_{,R}/f=-2$, we require that 
$0 \le m(r=-2) \le 1$ \cite{Faraoni,AGPT}.
The models proposed by Hu and Sawicki \cite{Hu07} and 
Starobinsky \cite{Star07} satisfy all these requirements.
They take the asymptotic form, $f(R) \simeq R-\mu R_c
[1-(R/R_c)^{-2n}]$ ($\mu>0, R_c>0, n>0$), in the region 
$R \gg R_c$ ($R_c$ is roughly the same order as $R_0$).
See Refs.~\cite{Li,AT,Appleby,Tsuji08,NO07} for 
other viable $f(R)$ models.

The main reason why viable $f(R)$ models are so restrictive
is that the strength of a coupling $Q$ between dark energy and non-relativistic 
matter (such as dark matter) is large 
in the Einstein frame ($Q=-1/\sqrt{6}$) \cite{APT}.
In the region of high-density where local gravity experiments are
carried out, the scalaron field $\phi$ needs to be almost frozen \cite{Nava,CT}
with a large mass through a chameleon mechanism \cite{KW}
to avoid that the field mediates a long ranged fifth force.
Cosmologically this means that the field does not approach a 
kinematically driven $\phi$ matter-dominated era ($``\phi$MDE'' \cite{coupled})
in which the evolution of scale factor is non-standard 
($a \propto t^{1/2}$ \cite{APT}).
The deviation from the $\Lambda$CDM model becomes 
important as the mass of the scalaron gets smaller so that 
the field begins to evolve slowly along its potential.
In other words the effect of modified gravity manifests itself
from the late-time matter era to the accelerated epoch \cite{Star07,Hu07}.
This leaves a number of interesting observational signatures for 
the equation of state of DE \cite{AT,Tsuji08}, 
matter power spectra \cite{staper,Star07,Tsuji08} and 
convergence spectra in weak lensing \cite{TT,Schmidt}.

One can generalize the analysis in $f(R)$ gravity to the theories
that have arbitrary constant couplings $Q$ \cite{TUMTY}.
In fact this is equivalent to Brans-Dicke theory \cite{BD}
with a scalar-field potential $V(\phi)$. 
By designing the potential so that the field 
mass is sufficiently heavy in the region of high density, it is 
possible to satisfy both local gravity and cosmological constraints
even when $|Q|$ is of the order of unity \cite{TUMTY}.
The representative potential of this type is given by 
$V(\phi)=V_0[1-C(1-e^{-2Q\phi})^p]$ $(V_0>0, C>0, 0<p<1)$, 
which covers the $f(R)$ models of Hu and 
Sawicki \cite{Hu07} and Starobinsky \cite{Star07} 
as special cases.
Especially when $|Q|$ is of the order of unity, these models lead to 
the large growth of matter density perturbations 
($\delta \propto t^{(\sqrt{25+48Q^2}-1)/6}$) at a late epoch 
of the matter era compared to the standard growth 
($\delta \propto t^{2/3}$) at an early epoch.
This gives rise to a significant change of the spectral index of
the matter power spectrum relative to that in the $\Lambda$CDM 
model \cite{Star07,Tsuji08}. 
Moreover it was recently shown that the convergence power 
spectrum in weak lensing observations is subject to a large 
modification by the non-standard evolution of 
matter perturbations \cite{TT}.

In this paper we shall study another test of modified gravity DE models
mentioned above by evaluating a normalized skewness,
$S_3=\langle \delta^3 \rangle/\langle \delta^2 \rangle^2$,
of matter perturbations.
The skewness provides a good test for the picture of 
gravitational instability from Gaussian initial conditions \cite{Ber}.
If large-scale structure grows via gravitational 
instability from Gaussian initial perturbations, the skewness 
in a Universe dominated by a pressureless matter
is known to be $S_3=34/7$ in General Relativity \cite{Peebles}.
Even when cosmological constant is present at late times, 
the skewness depends weakly on the expansion history of the Universe 
(less than a few percent) \cite{Marc,Benabed,Kofman}.
This situation hardly changes in open/closed universes \cite{curved}
and Dvali-Gabadadze-Poratti braneworld models \cite{Mul}.
One can see some difference for the models that are significantly 
different from Einstein gravity--such as Cardassian cosmologies \cite{Freese,Mul},
modified gravity models that respect Birkhoff's law \cite{Lue}.
In the context of dark energy coupled with dark matter, 
it was shown in Ref.~\cite{Luca04} that the skewness can be 
a probe of the violation of equivalence principle between 
dark matter and (uncoupled) baryons.

In Brans-Dicke theory with cosmological constant $\Lambda$
the skewness has been calculated in Ref.~\cite{BD2}
under the condition that the Brans-Dicke field is massless.
In this case the evolution of scale factor during the matter-dominated
epoch is given by $a(t) \propto t^{(2\omega_{\rm BD}+2)/
(3\omega_{\rm BD}+4)}$ \cite{BD2},
where $t$ is a cosmic time and $\omega_{\rm BD}$ is 
a Brans-Dicke parameter. 
If the field is massless, the Brans-Dicke parameter is constrained to be 
$\omega_{\rm BD}>40000$ \cite{Hoyle} from solar-system experiments.
This shows that the evolution of the scale factor in the matter 
era is very close to the standard one: $a(t) \propto t^{2/3}$.
We note that an effective gravitational ``constant'' that appears as 
a coefficient of matter density perturbations is also subject to change 
in Brans-Dicke theory. 
However it was found that the skewness in such a case
is given by $S_3=(34\omega_{\rm BD}+56)/
(7\omega_{\rm BD}+12)$ \cite{BD2} during the matter era,
which is very close to the standard one ($S_3=34/7$)
under the condition $\omega_{\rm BD}>40000$.

The $f(R)$ gravity corresponds to theory with the Brans-Dicke parameter 
$\omega_{\rm BD}=0$ \cite{Chiba}. 
Even in this situation, if the scalaron field has a potential whose mass is 
sufficiently large in the region of high density, the $f(R)$ models can pass local 
gravity constraints as in the models proposed in Refs.~\cite{Hu07,Star07}.
In such cases, compared to Brans-Dicke theory with a massless field, 
it is expected that the skewness may show significant  
deviations from that in General Relativity.
Since the evolution of scale factor and matter perturbations 
is different from that in the massless case, 
we can not employ the result of the skewness presented above.

In this paper we study second-order perturbations and the skewness 
for Brans-Dicke theory in the presence of a potential $V(\phi)$. 
This is equivalent to the scalar-field action given in Eq.~(\ref{action}) 
by identifying the coupling $Q$ with the Brans-Dicke 
parameter $\omega_{\rm BD}$
via the relation $1/(2Q^2)=3+2\omega_{\rm BD}$.
In the massless case the solar-system constraint, $\omega_{\rm BD}>40000$,
gives the bound $|Q| \lesssim 10^{-3}$,
but it is difficult to find some deviations from General Relativity
in such a situation.
Our interest is the case in which the coupling $Q$ is of the order of 
$0.1 \lesssim |Q| \lesssim 1$ with the field potential that has 
a sufficiently large mass in the high-density region.
This analysis includes viable $f(R)$ models \cite{Hu07,Star07}
recently proposed in the literature.
We would like to investigate how much extent the skewness
differs from that in the $\Lambda$CDM model.
We also derive conditions under which the contribution coming from
first-order field perturbations can be neglected relative to 
second-order matter and velocity perturbations by starting from
fully relativistic second-order perturbation equations.

%%%%%%%%%%%%%%%%%%%%%%%%%%
\section{Modified gravity models}
\label{back}
%%%%%%%%%%%%%%%%%%%%%%%%%%

The action for Brans-Dicke theory \cite{BD} in the presence of 
a potential $V$ is given by 
\begin{eqnarray}
\label{action0}
S = \int {\rm d}^4 x\sqrt{-g} \left[ \frac12 \chi R
-\frac{\omega_{\rm BD}}{2\chi} (\nabla \chi)^2 -V(\chi) 
\right]+S_m (g_{\mu \nu}, \Psi_m)\,,
\end{eqnarray}
where $\chi$ is a scalar field coupled to a Ricci 
scalar $R$, $\omega_{\rm BD}$ is a so-called Brans-Dicke 
parameter and $S_m$ is a matter action that depends on the
metric $g_{\mu \nu}$ and matter fields $\Psi_m$.
We shall use the unit $8\pi G=1$, but we restore 
the bare gravitational constant $G$ when it is required.

The action (\ref{action0}) is equivalent to the following 
scalar-tensor action with the correspondence $\chi=e^{-2Q\phi}$:
\begin{eqnarray}
\label{action}
S =  \int {\rm d}^4 x\sqrt{-g} \Bigg[ \frac12 F(\phi) R
-\frac12 \omega (\phi)(\nabla \phi)^2 -V(\phi) \Bigg]
+S_m (g_{\mu \nu}, \Psi_m)\,,
\end{eqnarray}
where 
\begin{eqnarray}
F(\phi)=e^{-2Q \phi}\,, \quad
\omega (\phi)=(1-6Q^2)F(\phi)\,.
\end{eqnarray}
As we already mentioned, the constant $Q$ is related with $\omega_{\rm BD}$
via the relation $1/(2Q^2)=3+2\omega_{\rm BD}$. 
In the limit $Q \to 0$ (i.e., $\omega_{\rm BD} \to \infty$),
the action (\ref{action}) reduces to the one for a minimally 
coupled scalar field $\phi$ with a potential $V(\phi)$.
The $f(R)$ gravity corresponds to the coupling $Q=-1\sqrt{6}$,
i.e., $\omega_{\rm BD}=0$.

In the absence of the potential $V(\phi)$
the coupling $Q$ is constrained to be $|Q| \lesssim 10^{-3}$
from solar-system tests.
We are interested in the case where the presence of the 
potential can make the models be consistent with 
local gravity constraints (LGC) even for $|Q|={\cal O}(1)$.
The representative potential of this type is given by \cite{TUMTY} 
\begin{eqnarray}
\label{potential}
V(\phi)=V_0 \left[ 1-C (1-e^{-2Q \phi})^p \right]~~~~~~
(V_0>0, C>0, 0<p<1)\,,
\end{eqnarray}
where $V_0$ is of the order of the present cosmological Ricci 
scalar $R_0$ in order to be responsible for the acceleration of 
the Universe today.
Note that the $f(R)$ models proposed by Hu and Sawicki \cite{Hu07}
and Starobinsky \cite{Star07} take the form $f(R)=R-\mu R_c [1-(R/R_c)^{-2n}]$ 
($\mu>0, R_c>0, n>0$) in the region $R \gg R_c$.
These $f(R)$ models are covered in the action (\ref{action}) 
with (\ref{potential}) by identifying the field potential 
to be $V=(RF-f)/2$ with $F=\partial f/\partial R=e^{2\phi/\sqrt{6}}$.

The background cosmological dynamics and LGC for the potential 
(\ref{potential}) have been discussed in details in Ref.~\cite{TUMTY}.
We review how the matter-dominated era is followed by 
the stage of a late-time acceleration. This is important when we discuss
the evolution of matter density perturbations in Sec.~\ref{skewness}.
As a matter source we take into account a non-relativistic 
matter with an energy density $\rho_m$.
In the flat Friedmann-Lemaitre-Robertson-Walker (FLRW) metric
with scale factor $a(t)$, where $t$ is cosmic time, 
the evolution equations for the action (\ref{action}) are
\begin{eqnarray}
\label{be1}
& & 3FH^2=\frac12 \omega \dot{\phi}^2+V
-3H\dot{F}+\rho_{m}\,, \\
\label{be2}
& & 2F\dot{H}=-\omega \dot{\phi}^2
-\ddot{F}+H\dot{F}-\rho_{m}\,, \\
\label{be3}
& & \omega \left( \ddot{\phi}+3H\dot{\phi}
+\frac{\dot{F}}{2F} \dot{\phi} \right)
+V_{,\phi}-\frac12 F_{,\phi}R=0\,,\\
\label{be4}
& & \dot{\rho}_{m}+3H \rho_{m}=0\,,
\end{eqnarray}
where $H\equiv\dot{a}/a$ is the Hubble parameter
and a dot represents a derivative with respect to $t$.
Note that the Ricci scalar is given by $R=6(2H^2+\dot{H})$.

We introduce the following dimensionless quantities:
\begin{eqnarray}
x_1 \equiv \frac{\dot{\phi}}{\sqrt{6}H}\,, \quad
x_2 \equiv \frac{1}{H}\sqrt{\frac{V}{3F}}\,,
\end{eqnarray}
and 
\begin{eqnarray}
\Omega_m \equiv \frac{\rho_m}{3FH^2}=
1-(1-6Q^2)x_1^2-x_2^2-2\sqrt{6}Q x_1\,,
\end{eqnarray}
where we used Eq.~(\ref{be1}). We then obtain
\begin{eqnarray}
\label{au1}
\frac{{\rm d} x_1}{{\rm d} N}
&=& \frac{\sqrt{6}}{2} (\lambda x_2^2-\sqrt{6} x_1)
+\frac{\sqrt{6}Q}{2} \left[ (5-6Q^2) x_1^2+
2 \sqrt{6}Q x_1-3x_2^2-1 \right]
-x_1 \frac{\dot{H}}{H^2}  \,, \\
\label{au2}
\frac{{\rm d} x_2}{{\rm d} N}
&=& \frac{\sqrt{6}}{2} (2Q-\lambda)x_1 x_2
-x_2 \frac{\dot{H}}{H^2} \,, 
\end{eqnarray}
where $N\equiv{\rm ln}\,(a)$ and 
$\lambda=-V_{,\phi}/V$.
The effective equation of state is defined by 
\begin{eqnarray}
w_{\rm eff}=-1-\frac23 \frac{\dot{H}}{H^2}\,,
\end{eqnarray}
where
\begin{eqnarray}
\frac{\dot{H}}{H^2}= -\frac{1-6Q^2}{2} \left[
3+3x_1^2-3x_2^2-6Q^2 x_1^2
+2\sqrt{6}Q x_1 \right]+
3Q (\lambda x_2^2-4Q)\,.
\end{eqnarray}

When $\lambda$ is a constant (i.e., $V(\phi)=V_0 e^{-\lambda \phi}$),
the fixed points of the system can be derived by setting 
${\rm d}x_1/{\rm d}N={\rm d}x_2/{\rm d}N=0$.
Even if $\lambda$ changes with time, as it is the case for the potential 
(\ref{potential}), the fixed points can be regarded
as instantaneous ones.
The following points can play the role of the matter-dominated epoch:
\begin{itemize}
\item (M1) $\phi$ matter-dominated era
\begin{eqnarray}
\label{fp1}
(x_1,x_2)=\left( \frac{\sqrt{6}Q}{3(2Q^2-1)}, 0 \right)\,,
\quad \Omega_{m}=\frac{3-2Q^2}{3(1-2Q^2)^2}\,,
\quad w_{\rm eff}=\frac{4Q^2}{3(1-2Q^2)}\,.
\end{eqnarray}
\item (M2) ``Instantaneous'' scaling solution
\begin{eqnarray}
\label{scaling}
(x_1,x_2)=\left( \frac{\sqrt{6}}{2\lambda},
\left[ \frac{3+2Q\lambda -6Q^2}{2\lambda^2} \right]^{1/2}
\right)\,,
\quad \Omega_{m}=1-\frac{3-12Q^2+7Q\lambda}
{\lambda^2}\,,
\quad w_{\rm eff}=-\frac{2Q}{\lambda}\,.
\end{eqnarray}
\end{itemize}
In order to realize the matter era ($\Omega_m \simeq 1$ and $w_{\rm eff} \simeq 0$)
by the point (M1), we require the condition $Q^2 \ll 1$.
This point was used in the coupled quintessence scenario \cite{coupled}
(in the Einstein frame) where the coupling is 
constrained to be $|Q| \lesssim 0.1$ from
Cosmic Microwave Background anisotropies.
In $f(R)$ gravity ($Q=-1/\sqrt{6}$) we have $\Omega_m=2$ 
and $w_{\rm eff}=1/3$ (i.e., $a \propto t^{1/2}$ \cite{APT}), 
which means that the point (M1) can not be responsible for   
the matter era for $|Q|$ of the order of unity.

The matter era can be realized by the point (M2)
for $|\lambda| \gg |Q|={\cal O}(1)$.
The parameter $\lambda$ for the potential (\ref{potential}) 
is given by 
\begin{eqnarray}
\lambda=\frac{2CpQe^{-2Q \phi} (1-e^{-2Q \phi})^{p-1}}
{1-C(1-e^{-2Q \phi})^p}\,,
\end{eqnarray}
which is much larger than 1 for $|Q \phi | \ll 1$ (provided that 
$C$ and $p$ are not very much smaller than 1).
Since $R \simeq \rho_{m}/F$ during the deep 
matter-dominated epoch, the field $\phi$ is stuck at the 
instantaneous minima characterized by the condition
$V_{,\phi} (\phi_m)+Q\rho_{m} \simeq 0$ [see Eq.~(\ref{be3})].
For the potential (\ref{potential}) this translates into
\begin{eqnarray}
\label{phim}
\phi_m \simeq \frac{1}{2Q} \left(
\frac{2V_0pC}{\rho_m} \right)^{\frac{1}{1-p}}\,,
\end{eqnarray}
which means that $|Q \phi_m | \ll 1$ and hence 
$|\lambda| \gg 1$ during the deep matter era ($\rho_m \gg V_0$).
When $|Q|={\cal O}(1)$, the matter era is realized by 
the point (M2) instead of (M1).

For the dynamical system given by Eqs.~(\ref{au1}) and (\ref{au2})
there exist the following fixed points 
that lead to the late-time acceleration:
\begin{itemize}
\item (A1) Scalar-field dominated point
\begin{eqnarray}
\label{sp1}
(x_1,x_2)=\left( \frac{\sqrt{6}(4Q-\lambda)}
{6(4Q^2-Q\lambda-1)}, \left[ \frac{6-\lambda^2+8Q\lambda
-16Q^2}{6(4Q^2-Q\lambda-1)^2} \right]^{1/2} \right)\,,
~~\Omega_{m}=0\,,
\quad w_{\rm eff}=-\frac{20Q^2-9Q\lambda-3+\lambda^2}
{3(4Q^2-Q\lambda -1)}.
\end{eqnarray}
\item (A2) de Sitter point (present for $\lambda=4Q$)
\begin{eqnarray}
(x_1,x_2)=(0,1)\,,
\quad \Omega_{m}=0\,,
\quad w_{\rm eff}=-1\,.
\end{eqnarray}
\end{itemize}
The de Sitter point (A2) appears only in the presence 
of the coupling $Q$ (characterized by the condition 
$V_{,\phi}+QFR=0$ in Eq.~(\ref{be3}), i.e., $\lambda=4Q$).
This can be regarded as the special case of 
the accelerated point (A1).
For the potential (\ref{potential}) the parameter $|\lambda|$
is much larger than $|Q|$ during the matter era, but it gradually 
becomes the same order as $|Q|$ as the system enters the 
accelerated epoch.
It was shown in Ref.~\cite{TUMTY} that the de Sitter point (A2)
is stable for ${\rm d} \lambda/{\rm d}\phi<0$.
As long as $|\lambda|$ continues to decrease with the 
growth of $|\phi|$, the solutions are finally trapped 
at the stable de Sitter point (A2).
If the stability condition, ${\rm d} \lambda/{\rm d}\phi<0$, 
is not satisfied, the solutions approach another accelerated 
point (A1).

In the following we are mainly interested in the case where 
the ``instantaneous'' matter point (M2) is followed by the 
de Sitter point (A2).
During most stages of cosmic expansion history
the field $\phi$ is trapped at instantaneous 
minima of an effective potential induced by the matter coupling.
This means that the condition, $\dot{\phi}^2 \ll H^2$, 
is well satisfied.

The mass squared of the field $\phi$
for the potential (\ref{potential}) is given by 
\begin{eqnarray}
\label{mass}
M^2 \equiv V_{,\phi \phi}=4V_0CpQ^2 e^{-2Q \phi}
(1-e^{-2Q \phi})^{p-2} (1-p e^{-2 Q\phi})\,,
\end{eqnarray}
which is much larger than $R_0$ $(\sim V_0)$ in the region $R \gg R_0$.
In this situation it is possible to satisfy local gravity constraints
in the region of high density \cite{Hu07,Nava,CT} 
through a chameleon mechanism \cite{KW}.
Since the field is massive inside a spherical symmetric body with radius 
$r_c$, only the surface part of its mass distribution contributes 
to the field profile outside the body. 
The effective coupling $Q_{\rm eff}$ between 
the field and the pressureless matter is suppressed by a thin-shell parameter
$\Delta r_c/r_c$ relative to the bare coupling $Q$.
For the potential (\ref{potential}) it was shown in Ref.~\cite{CT}
that constraints coming from 
solar system tests as well as the violation of equivalence
principle give the bounds $p>1-5/(9.6-{\rm log}_{10}|Q|)$ and 
$p>1-5/(13.8-{\rm log}_{10}|Q|)$, respectively.
In $f(R)$ gravity these constraints correspond to 
$p>0.50$ and $p>0.65$, respectively.

Substituting the field value (\ref{phim}) for Eq.~(\ref{mass}), we find that 
the mass squared during the matter era is given by 
\begin{eqnarray}
M^2 \simeq \left( \frac{3^{2-p}}{2^p pC} \right)^{\frac{1}{1-p}}
(1-p)Q^2 \left( \frac{H^2}{V_0} \right)^{\frac{1}{1-p}}H^2\,,
\end{eqnarray}
where we used $3H^2 \simeq \rho_m$.
We then find that the inequality, $M^2 \gg H^2$, holds 
for the values of $p, C, Q$ not very much smaller than unity.
In the next section we shall use this property when we derive the equation 
for matter perturbations approximately. 

%%%%%%%%%%%%%%%%%%%%%%%%%%
\section{Second-order cosmological perturbations}
\label{secmodel}
%%%%%%%%%%%%%%%%%%%%%%%%%%

In this section we consider second-order cosmological perturbations 
for the action (\ref{action}) and derive the equation for 
matter perturbations approximately.

\subsection{Perturbation equations}

Let us start with a perturbed metric including scalar metric 
perturbations $\alpha$, $\beta$, $\varphi$ and $\gamma$ 
about the flat FLRW background \cite{metper}:
\begin{eqnarray}
\rd s^2=-(1+2\alpha) \rd t^2-2a \beta_{,i}
\rd t \rd x^{i}+a(t)^2 \left[(1+2\varphi)
\delta_{ij}+2\gamma_{|ij} \right] \rd x^i \rd x^j\,.
\end{eqnarray}
At the second-order the scalar variables are written as 
\begin{eqnarray}
\alpha \equiv \alpha^{(1)}+\alpha^{(2)}\,,~~~
\beta \equiv \beta^{(1)}+\beta^{(2)}\,,~~~
\varphi \equiv \varphi^{(1)}+\varphi^{(2)}\,,~~~
\gamma \equiv \gamma^{(1)}+\gamma^{(2)}\,,
\end{eqnarray}
where the subscripts represent the orders of perturbations.
We introduce the following quantities:
\begin{eqnarray}
\label{chi}
\chi \equiv a(\beta+a\dot{\gamma})\,,\quad
\kappa \equiv \delta K\,,
\end{eqnarray}
where $\delta K$ is the perturbation of an extrinsic
curvature $K$.

We decompose the scalar field $\phi$ and the 
quantity $F$ into background and perturbed parts:
\begin{eqnarray}
\phi=\phi_0(t)+\delta \phi (t,\bm{x})\,,
\quad
F=F_0(t)+\delta F (t,\bm{x})\,,
\end{eqnarray}
where $\delta \phi$ and $\delta F$
depend on $t$ and a position vector $\bm{x}$.
In the following we omit the subscript ``0'' 
from background quantities.
The components of the energy momentum tensor of 
a pressureless matter can be decomposed as
\begin{eqnarray}
\label{Ten}
T^0_0=-(\rho_m+\delta \rho_m)\,,\quad
T^0_i=-\rho_m v_{,i} \equiv q_{,i}\,,
\end{eqnarray}
where $v$ is a rotational-free velocity potential.
At the second-order, the perturbed quantities can be 
explicitly written as
\begin{eqnarray}
\delta \phi \equiv \delta \phi^{(1)}+\delta \phi^{(2)}\,,~~~
\delta F \equiv \delta F^{(1)}+\delta F^{(2)}\,,~~~
\delta \rho_m \equiv \delta \rho_m^{(1)}+\delta \rho_m^{(2)}\,,~~~
v \equiv v^{(1)}+v^{(2)}\,.
\end{eqnarray}

The perturbation equations for the action (\ref{action}), 
up to the second-order, have been derived in Ref.~\cite{Noh}
(see also Refs.~\cite{HN}).
They are given by 
\begin{eqnarray}
\label{per1}
& & \kappa-3H\alpha+3\dot{\varphi}+\frac{\Delta}{a^2}\chi
=-\alpha \left(\frac92 H\alpha-\frac{1}{a}{\beta^{,i}}_{|i} \right)+\frac32 H
\beta^{,i} \beta_{,i}\,, \\
\label{per2}
& & 4\pi G \delta \rho_{\rm eff}+H\kappa+\frac{\Delta}{a^2}\varphi
=\frac16 \kappa^2-\frac{1}{4a^2} \beta_{(,i|j)} \beta^{,i|j}
+\frac{1}{12a^2} ({\beta^{,i}}_{|i})^2\,,\\
\label{per3}
& & \kappa+\frac{\Delta}{a^2}\chi-12\pi G \rho av-
\frac{3}{2F} \left( \omega \dot{\phi} \delta \phi+
\delta \dot{F}-H \delta F -\dot{F} \alpha \right) \nonumber \\
& &=\Delta^{-1} \nabla^i \left[ -\alpha (\kappa_{,i}+12\pi G a q_{,i})
+\frac{3}{4a} \alpha_{,j} ({\beta^{,j}}_{|i}+{\beta_{,i}}^{|j})
-\frac{1}{2a}\alpha_{,i} {\beta^{,j}}_{|j} \right]\,,\\
\label{per4}
& & \dot{\kappa}+2H\kappa-4\pi G (\delta \rho_{\rm eff}
+3\delta P_{\rm eff})+\left( 3\dot{H}+\frac{\Delta}{a^2} \right) \alpha
\nonumber \\
& &=\alpha \dot{\kappa}-\frac{1}{a}\kappa_{,i}\beta^{,i}
+\frac13 \kappa^2+\frac32\dot{H} (\alpha^2-\beta^{,i}\beta_{,i})
+\frac{1}{a^2}\left( 2\alpha {\alpha^{|i}}_i+\alpha_{,i}\alpha^{,i}
-\beta^{,j} {{\beta_{,j}}^{|i}}_i -\beta^{,j|i} \beta_{,j|i}\right)
+\frac{1}{a^2}\beta_{(,i|j)}\beta^{,i|j}
-\frac{1}{3a^2} ({\beta^{,i}}_{|i})^2, \nonumber \\ \\
\label{per5}
& & 
\delta \dot{\rho}_m+3H \delta \rho_m-
\rho_m \left(\kappa-3H \alpha+\frac{1}{a} \Delta v \right) \nonumber \\
& & =-\frac{1}{a} \delta \rho_{m,i} \beta^{,i}
+\delta \rho_m (\kappa-3H\alpha)
+\rho_m \left[ \alpha \kappa+\frac32 H (\alpha^2
-\beta^{,i} \beta_{,i}) \right]
-\frac{1}{a} \left( \alpha {q^i}_{|i}+2q^i \alpha_{,i} \right)\,,\\
\label{per6}
& & \dot{v}+Hv-\frac{1}{a}\alpha
=-\frac{1}{\rho_m} \Delta^{-1} \nabla^i
\left[q_{,i} (\kappa-3H\alpha)+\frac{1}{a}
\left\{ -q_{,i|j}\beta^{,j}-q_{,j} {\beta^{,j}}_{|i}
-\delta \rho_m \alpha_{,i}
+\rho_m (\alpha \alpha_{,i}-\beta^{,j} \beta_{,j|i}) \right\}
\right]\,,
\end{eqnarray}
where 
\begin{eqnarray}
\label{delrho}
\delta \rho_{\rm eff} &=&
\frac{1}{8\pi G F} \biggl[ \delta \rho_m+\omega 
(\dot{\phi} \delta \dot{\phi}-\alpha \dot{\phi}^2)+
\frac12 \omega_{,\phi} \delta \phi \dot{\phi}^2
-\frac12 (F_{,\phi}R-2V_{,\phi})\delta \phi-3H \delta \dot{F}
+\biggl( \frac12 R+\frac{\Delta}{a^2} \biggr)\delta F \nonumber \\
&&~~~~~~~~~
+\left(6H \alpha-\frac{\Delta}{a^2}\chi -3\dot{\varphi}
\right) \dot{F}-\frac{\delta F}{F}
\left( \rho_m+\frac12 \omega \dot{\phi}^2
+V-3H\dot{F} \right) \biggr]\,,\\
\label{delP}
\delta P_{\rm eff} &=&
\frac{1}{8\pi G F} \biggl[ \omega 
(\dot{\phi} \delta \dot{\phi}-\alpha \dot{\phi}^2)
+\frac12 \omega_{,\phi} \delta \phi \dot{\phi}^2
+\frac12 (F_{,\phi}R-2V_{,\phi})\delta \phi
+\delta \ddot{F}+2H\delta \dot{F}
-\biggl( \frac12 R+\frac23\frac{\Delta}{a^2} \biggr)\delta F
\nonumber \\
&&~~~~~~~~~
-2\alpha \ddot{F}
-\left(\dot{\alpha}+4H\alpha-\frac23 \frac{\Delta}{a^2}\chi
-2\dot{\varphi} \right) \dot{F}
-\frac{\delta F}{F} \left( \frac12 \omega \dot{\phi}^2
-V+\ddot{F}+2H\dot{F} \right) \biggr]\,.
\end{eqnarray}

The equations for the perturbations $\delta \phi$ and $\delta F$ are  
\begin{eqnarray}
\label{delphi}
& & \delta \ddot{\phi}+\left(3H+\frac{\omega_{,\phi}}{\omega}
\dot{\phi} \right)\delta \dot{\phi}
+\left[ -\frac{\Delta}{a^2}+\left( \frac{\omega_{,\phi}}{\omega}
\right)_{,\phi} \frac{\dot{\phi}^2}{2}+
\left( \frac{2V_{,\phi}-F_{,\phi}R}{2\omega} \right)_{,\phi} \right]
\delta \phi-\dot{\phi}\dot{\alpha}
-\left(2\ddot{\phi}+3H\dot{\phi}+\frac{\omega_{,\phi}}{\omega}
\dot{\phi}^2 \right)\alpha \nonumber \\
& &-\dot{\phi}\kappa-\frac{1}{2\omega} F_{,\phi} \delta R=N_{\delta \phi}\,,\\
\label{delF}
& & \delta \ddot{F}+3H \delta \dot{F}+\left(-\frac{\Delta}{a^2}
-\frac{R}{3} \right) \delta F+\frac23 \omega \dot{\phi} \delta \dot{\phi}
+\frac13 (\omega_{,\phi} \dot{\phi}^2+2F_{,\phi}R-4V_{,\phi})\delta \phi
-\frac13 \delta \rho_m -\dot{F} (\kappa+\dot{\alpha}) \nonumber \\
& &-\left(\frac23 \omega \dot{\phi}^2+2\ddot{F}+3H \dot{F} \right)\alpha
+\frac13 F \delta R=N_{\delta F}\,,
\end{eqnarray}
where $N_{\delta \phi}$ and $N_{\delta F}$ are  
second-order terms whose explicit expressions
are given in Ref.~\cite{Noh}.

At the first-order the quantity, 
$\delta \rho^{(1)} \equiv \delta \rho_m^{(1)}
-\dot{\rho}_m av^{(1)}$, is known to be gauge-invariant \cite{HN}.
In order to construct gauge-invariant variables 
at the second-order, we introduce the following quantities
\begin{eqnarray}
\label{ga1}
& &\delta \rho \equiv \delta \rho_m-\dot{\rho}_m av
+\delta \rho^{(q)}\,,\\
\label{ga2}
& &v_\chi \equiv v-\frac{1}{a}\chi+v_{\chi}^{(q)}\,,
\end{eqnarray}
where $\delta \rho^{(q)}$ and $v_{\chi}^{(q)}$ 
are quadratic combinations of first-order terms.
By defining $\delta_m \equiv \delta \rho_m/\rho_m$, 
it was shown in Ref.~\cite{Noh} that the following quantity 
is gauge-invariant at the second-order:
\begin{eqnarray}
\label{delmgau}
\delta \equiv \delta_m+3aHv
-\frac{\delta \dot{\rho}_v}{\rho_m} av
+\frac32 \rho_m \dot{H}a^2v^2
-v^{,i}v_{,i}-3\rho_m aH \Delta^{-1} \nabla^{i}
\left( \frac{\delta \rho_v}{\rho_m}v_{,i} \right)\,,
\end{eqnarray}
where $\delta \rho_v \equiv\delta \rho_m
-\dot{\rho}_m av$.
Note that the quantity $v_\chi$ can be also made 
gauge-invariant \cite{Noh}.

\subsection{Approximate second-order equations}

If we take the temporal comoving gauge ($v=0$),
we have $\delta=\delta_m$
and $q_{,i}=0$ [see Eqs.~(\ref{Ten}) and (\ref{delmgau})]. 
Taking $\gamma=0$ for the spatial
gauge condition, it follows that $\beta=\chi/a$ from 
Eq.~(\ref{chi}).
{}From Eq.~(\ref{per6}) we obtain
\begin{eqnarray}
\alpha=-\frac12 \beta_{,i}\beta^{,i}\,,
\end{eqnarray}
which means that $\alpha$ is a second-order quantity.
Up to the second-order, Eqs.~(\ref{per3}) and (\ref{per5}) are
written as
\begin{eqnarray}
\label{kappaes}
& &\kappa=-\frac{\Delta}{a^2} \chi
+\frac{3}{2F} (\omega \dot{\phi} \delta \phi
+\delta \dot{F}-H \delta F-\dot{F} \alpha)\,, \\
\label{dotdelta}
& &\dot{\delta}-\kappa=\kappa \delta
-\frac{1}{a} \delta_{,i} \beta^{,i}\,.
\end{eqnarray}

In order to evaluate the terms on the r.h.s. of Eq.~(\ref{dotdelta}), 
it is sufficient to consider Eqs.~(\ref{ga2}) and (\ref{kappaes})
at the first order with the gauge $v=0$.
We then have $\chi^{(1)}=-av_\chi$, $\beta^{(1)}=-v_\chi$
and 
\begin{eqnarray}
\label{kappa1}
\kappa^{(1)}=\frac{\Delta v_\chi}{a}+\frac{3}{2F}
(\omega \dot{\phi} \delta \phi
+\delta \dot{F}-H \delta F)\,,
\end{eqnarray}
where we omitted the order of the subscript  from the r.h.s. of 
these equations.
Hence Eq.~(\ref{dotdelta}) can be read as
\begin{eqnarray}
\label{deleq}
\dot{\delta}-\kappa=\frac{1}{a}
\nabla \cdot (\delta \nabla v_{\chi})+
\frac{3}{2F} (\omega \dot{\phi} \delta \phi
+\delta \dot{F}-H\delta F)\delta\,.
\end{eqnarray}
In the following we consider a situation in which
the scalar-field dependent terms on the r.h.s. of 
Eq.~(\ref{kappa1}) is neglected relative to the term 
$\Delta v_\chi/a$.
In this case Eq.~(\ref{deleq}) yields
\begin{eqnarray}
\label{deleq2}
\dot{\delta}-\kappa \simeq \frac{1}{a}
\nabla \cdot (\delta \nabla v_{\chi})\,.
\end{eqnarray}
Later we shall confirm the validity of this approximation.

{}From Eq.~(\ref{per4}) with Eqs.~(\ref{per1}), (\ref{delrho}) 
and (\ref{delP}) we obtain
\begin{eqnarray}
\label{kappa}
& &\dot{\kappa}+\left(2H +\frac{\dot{F}}{2F} \right) \kappa
-\frac{1}{2F} \left[ \delta \rho +4\omega \dot{\phi} 
\delta \dot{\phi}+(2\omega_{,\phi} \dot{\phi}^2
+F_{,\phi}R-2V_{,\phi})\delta \phi
+3\delta \ddot{F}+3H\delta \dot{F}-\left(6H^2
+\frac{\Delta}{a^2} \right)\delta F \right] \nonumber \\
& & =\frac{N_0}{2}\frac{\dot{F}}{F}+N_3
-\frac{3\dot{F}}{2F}\dot{\alpha}
-\left[ 3\dot{H}+\frac{1}{2F} (6\ddot{F}+6H\dot{F}
+4\omega \dot{\phi}^2)+\frac{\Delta}{a^2} \right]\alpha\,,
\end{eqnarray}
where $N_0$ and $N_3$ correspond to the second-order terms 
on the r.h.s. of Eqs.~(\ref{per1}) and (\ref{per4}), respectively.
Following Refs.~\cite{Boi,CST,TUMTY,Tsuji07} we employ the sub-horizon approximation 
under which the terms containing $\kappa$, $\delta \rho$, $\Delta\,\delta F/a^2$ and 
$\Delta \alpha/a^2$ are picked up in Eq.~(\ref{kappa}). 
Note that $|\dot{F}/HF| \ll 1$ under the condition $|\dot{\phi}| \ll H$.
Apart from the term $\Delta \alpha/a^2$, the terms on the r.h.s. 
of Eq.~(\ref{kappa}) are of the order of $H^2 \alpha$ or smaller.
We then have
\begin{eqnarray}
\label{kappa2}
\dot{\kappa}+2H\kappa
-\frac{1}{2F} \left( \delta \rho  
-\frac{\Delta}{a^2} \delta F \right) 
\simeq \frac{1}{a^2} \left[ (\nabla v_{\chi}) \cdot
(\nabla v_{\chi})^{,i} \right]_{,i}\,.
\end{eqnarray}
Of course this approximation is justified  when the second-order
term on the r.h.s. of Eq.~(\ref{kappa2}) is larger than the 
first-order (field-dependent) terms on the l.h.s. of 
Eq.~(\ref{kappa}) we have neglected.
Later we shall derive conditions 
under which this approximation is valid.

Let us estimate the field perturbation $\delta \phi$
as well as $\delta F$.
As we explained in the previous section,  
the field mass $M$ defined in Eq.~(\ref{mass})
is much larger than $H$.
Using the approximation in which the terms containing $M^2$,
$\Delta\, \delta \phi/a^2$, $\Delta\,\delta F/a^2$, 
$\delta \rho$ and $\delta R$ are dominant contributions to
Eqs.~(\ref{delphi}) and (\ref{delF}), we obtain
\begin{eqnarray}
\label{delphi2}
& & \left(-\frac{\Delta}{a^2}+\frac{M^2}{\omega} \right)
\delta \phi-\frac{\dot{\phi}}{a} \Delta v_\chi
-\frac{1}{2\omega}F_{,\phi} \delta R
 \simeq 0\,,\\
 \label{delF2}
& & -\frac{\Delta}{a^2}\delta F-\frac13 \delta \rho
-\frac{\dot{F}}{a}\Delta v_\chi
+\frac13 F \delta R \simeq 0\,.
\end{eqnarray}
This approximation is accurate as long as an oscillating mode of
the field perturbation does not dominate over 
the matter-induced mode \cite{Star07,TUMTY}.
Note that we have neglected second-order terms on the r.h.s. of 
Eqs.~(\ref{delphi}) and (\ref{delF}).
Since the field is nearly frozen at the instantaneous minimum 
given in Eq.~(\ref{phim}), the dominant second-order term 
corresponds to $V_{,\phi \phi \phi} \delta \phi^2$.
This term gives rise to only a tiny correction to the growth 
rate of perturbations. 
Moreover it can be neglected relative to the second-order 
term on the r.h.s. of Eq.~(\ref{kappa2}).
See Appendix for the detailed estimation of such a second-order term.

On combining Eqs.~(\ref{delphi2}) and (\ref{delF2}), we find
\begin{eqnarray}
\label{M2}
\left( \frac{M^2}{F}-\frac{\Delta}{a^2} \right) \delta F
=2Q^2 \delta \rho+\frac{\dot{F}}{a} \Delta v_\chi\,.
\end{eqnarray}
Note that $\delta \rho=\rho_m \delta \simeq 3FH^2 \delta$
during the matter era. 
At the first-order we also have the following relation from 
Eq.~(\ref{deleq2}):
\begin{eqnarray}
\label{delves}
\frac{1}{a} \Delta v_{\chi}
=\kappa=\dot{\delta}=
cH \delta\,,~~~~~c\equiv \dot{D}/HD\,,
\end{eqnarray}
where $D(t)$ is the time-dependent part of $\delta$.
Since $D(t)$ is typically proportional to $t^n$
with $n$ of the order of unity \cite{TUMTY},
it follows that $c={\cal O}(1)$.
Hence we get 
\begin{eqnarray}
\left|\frac{(\dot{F}/a)\Delta v_\chi}
{2Q^2 \delta \rho}
\right| \simeq
\left| \frac{\dot{\phi}}{QH} \right|\,. 
\end{eqnarray}
As long as the condition,
\begin{eqnarray}
\label{QH}
|\dot{\phi}| \ll |QH| \,,
\end{eqnarray}
is satisfied, we have that 
$|2Q^2 \delta \rho | \gg |(\dot{F}/a)\Delta v_\chi|$ and 
\begin{eqnarray}
\label{M3}
\left( \frac{M^2}{F}-\frac{\Delta}{a^2} \right) \delta F
\simeq 2Q^2 \rho_m \delta\,.
\end{eqnarray}
In the previous section we showed that $|\dot{\phi}|$
is much smaller than $H$ for the 
potential (\ref{potential}).
Hence the condition (\ref{QH}) holds well
for the values of $|Q|$ which are not very much
smaller than 1.

Equation (\ref{M3}) shows that the perturbation $\delta F$
is sourced by the matter perturbation $\delta$.
Hence Eq.~(\ref{kappa2}) can be written as 
\begin{eqnarray}
\label{kappa3}
\dot{\kappa}+2H \kappa
-4\pi \rho_m G_{\rm eff}  \delta
=\frac{1}{a^2} \left[ (\nabla v_{\chi}) \cdot
(\nabla v_{\chi})^{,i} \right]_{,i}\,,
\end{eqnarray}
where 
\begin{equation}
\label{Geff}
G_{\rm eff} \delta \equiv \frac{1}{8\pi F}
\left( \delta-\frac{1}{\rho_m}
\frac{\Delta \delta F}{a^2} \right)\,.
\end{equation}

We introduce an effective potential $\Phi$ and a peculiar 
velocity $\bm{u}$ as follows:
\begin{eqnarray}
\label{dePhi}
\frac{\Delta \Phi}{a^2} &=& 4\pi \rho_m G_{\rm eff}
\delta\,,\\
\bm{u} &=&-\nabla v_{\chi}\,.
\end{eqnarray}
If one defines an effective gravitational potential $\Psi=\varphi+aHv_{\chi}$
in Eq.~(\ref{per2}), it follows that $\Delta \Psi/a^2=
4\pi \tilde{G}_{\rm eff} \rho_m \delta$ at linear order,  
where $\tilde{G}_{\rm eff}$ is different from $G_{\rm eff}$
by the sign of $\Delta F/a^2$.
Taking the time-derivative of Eq.~(\ref{deleq2}) together with the 
use of Eq.~(\ref{kappa3}), we get 
\begin{eqnarray}
\label{basic3}
\frac{\partial^2 \delta}{\partial t^2}+2H \frac{\partial \delta}
{\partial t}=\frac{1}{a^2} \nabla \cdot (1+\delta) \nabla \Phi+
\frac{1}{a^2} \frac{\partial^2}{\partial x^i \partial x^j}
\left( u^i u^j \right)\,.
\end{eqnarray}
This is our master equation that is used to compute the 
skewness of matter density perturbations in the next section.

%%%%%%%%%%%%%%%%%%%%%%%%%%
\section{Skewness in modified gravity}
\label{skewness}
%%%%%%%%%%%%%%%%%%%%%%%%%%

In this section we study the skewness of matter perturbations
for the action (\ref{action}) with the potential (\ref{potential}).
The skewness will be derived analytically in the matter-dominated
epoch by using Eq.~(\ref{basic3}).

\subsection{First-order perturbations}

We write the solution to  Eq.~(\ref{basic3})
in the form $\delta=\delta^{(1)}+\delta^{(2)}+\cdots$, 
where the subscripts represent the orders of perturbations.
The equation for the first-order perturbation $\delta^{(1)}$ is
\begin{eqnarray}
\label{firstorder}
\frac{\partial^2 \delta^{(1)}}{\partial t^2}+
2H \frac{\partial \delta^{(1)}}{\partial t}-
4\pi \rho_m G_{\rm eff}^{(1)}\,\delta^{(1)}=0\,.
\end{eqnarray}
We express the first-order perturbations $\delta^{(1)}$ and $\delta F^{(1)}$
in the plane-wave form:
$\delta^{(1)}= \int \delta_k^{(1)}(t)e^{-i {\bm k} \cdot {\bm x}}\,{\rm d}^3{\bm k}$ and
$\delta F^{(1)}= \int \delta F_k^{(1)}(t)e^{-i {\bm k} \cdot {\bm x}}\,{\rm d}^3{\bm k}$.
{}From Eq.~(\ref{M3}) we obtain
\begin{eqnarray}
\label{delF3}
\delta F_k^{(1)} (t)=
\frac{2Q^2 \rho_m}
{M^2/F+k^2/a^2}\delta_k^{(1)}(t)\,.
\end{eqnarray}
Then the temporal part of Eq.~(\ref{firstorder}) satisfies
\begin{eqnarray}
\label{dotH}
\ddot{\delta}_k^{(1)} (t)+2H\dot{\delta}_k^{(1)} (t)
-4\pi \rho_m  G_{\rm eff}^{(1)} \delta_k^{(1)} (t)=0\,,
\end{eqnarray}
where 
\begin{eqnarray}
\label{Geff2}
G_{\rm eff}^{(1)}=\frac{1}{8\pi F}
\frac{(k^2/a^2)(1+2Q^2)+M^2/F}
{(k^2/a^2)+M^2/F}\,.
\end{eqnarray}

In the early stage of the matter era, the mass $M$
is sufficiently heavy to satisfy the condition $M^2/F \gg k^2/a^2$.
In this regime we have $G_{\rm eff}^{(1)} \simeq 
1/8\pi F \simeq G$, thus mimicking the evolution 
in General Relativity. 
At late times it happens that the perturbations
enter the regime $M^2/F \ll k^2/a^2$.
This case corresponds to $G_{\rm eff}^{(1)} \simeq 
(1+2Q^2)/8\pi F$, thus showing the deviation from 
General Relativity.
The transition from the former regime to the latter regime 
occurs at a redshift $z_k$ given by \cite{TUMTY}:
\begin{eqnarray}
\label{zk}
z_k \simeq \left[\left( \frac{k}{a_0H_0} \frac{1}{Q}
\right)^{2(1-p)} \frac{2^p pC}{(1-p)^{1-p}}
\frac{1}{(3F_0 \Omega_m^{(0)})^{2-p}}
\frac{V_0}{H_0^2} \right]^{\frac{1}{4-p}}-1\,,
\end{eqnarray}
where $a_0$ and $H_0$ are the present values.

Using the derivative with respect to 
$N={\rm ln}\,(a)$, Eq.~(\ref{dotH}) can be written as 
\begin{eqnarray}
\label{dotH2}
\delta_k^{(1)''}+\left( \frac12- \frac32 w_{\rm eff} \right)
\delta_k^{(1)'}-12\pi  F \Omega_m 
G_{\rm eff}^{(1)} \delta_k^{(1)}=0\,,
\end{eqnarray}
where a prime represents the derivative in terms of $N$.
As we explained in Sec.~\ref{back}, we are considering  
the case in which the matter era is realized
by the point (M2) with $|\lambda | \gg |Q|={\cal O}(1)$.
Since $\Omega_m \simeq 1$ and $w_{\rm eff} \simeq 0$
in this case, we get the following solutions  
\begin{equation}
\label{del1}
  \delta_k^{(1)}(t) \propto
  \begin{cases}
     t^{2/3}\,, & \text{for $z \gg z_k$,}\\
     t^{\frac16 (\sqrt{25+48Q^2}-1)}, & \text{for $z \ll z_k$.}
  \end{cases}  
\end{equation}
In these asymptotic regimes the growth rates of first-order 
perturbations are independent of the wavenumber $k$.
The growth rate is constrained to be 
$s \equiv \dot{\delta}^{(1)}_k/H\delta^{(1)}_k \lesssim 2$
from observational data, which gives the bound 
$|Q| \lesssim 1$ \cite{TUMTY}.

\subsection{Conditions for the validity of approximations
to reach Eq.~(\ref{basic3})}

In order to reach Eq.~(\ref{deleq2}), we have employed the 
approximation that the field-dependent term 
on the r.h.s of Eq.~(\ref{kappa1}) is neglected relative to
the term $\Delta v_\chi/a$.
We have also neglected some of first-order terms 
in Eq.~(\ref{kappa}) relative to the second-order term
$\frac{1}{a^2} \left[ (\nabla v_{\chi}) \cdot
(\nabla v_{\chi})^{,i} \right]_{,i}$ in Eq.~(\ref{kappa2}).
We derive conditions under which these approximations 
are justified. 

We write the temporal part of the first-order perturbation
$v_{\chi}^{(1)}$ as $(v_{\chi})_k^{(1)} (t)$.
Using Eqs.~(\ref{delF3}) and (\ref{delves}) 
together with the relation $\rho_m \simeq 3FH^2$
that holds during the matter era, we find
\begin{eqnarray}
\label{delFes}
\left| \frac{H \delta F_k^{(1)}}{F} \right| \simeq
\frac{Q^2H^2}{M^2/F+k^2/a^2}  
\left| H\delta_k^{(1)} \right|
\simeq
\frac{Q^2H^2}{M^2/F+k^2/a^2}  
\left|
\frac{1}{a} (\Delta v_\chi)_k^{(1)} \right|\,.
\end{eqnarray}
As we showed in Sec.~\ref{back}
the condition, $M^2/F \gg H^2$, holds for the potential 
(\ref{potential}).
Moreover we are considering sub-horizon modes deep inside the horizon,
i.e., $k^2 \gg a^2 H^2$.
This leads to the relation $|H \delta F_k^{(1)}/F| \ll |(\Delta v_\chi)_k^{(1)}/a|$
in Eq.~(\ref{delFes}).
Since the field $\phi$ is nearly frozen at instantaneous minima of its
effective potential, we have the relation 
$|\delta \dot{F}_k^{(1)}| \lesssim |H \delta F_k^{(1)}|$ and hence 
$|\delta \dot{F}_k^{(1)}/F| \ll |(\Delta v_\chi)_k^{(1)}/a|$.
The following inequality is also satisfied:
\begin{eqnarray}
\left| \frac{1}{F}\omega \dot{\phi} \delta \phi_k^{(1)} \right|
=\left| \frac{(1-6Q^2)\dot{\phi}}{2Q} 
\frac{\delta F_k^{(1)}}{F} \right| \ll
\left| \frac{1}{a} (\Delta v_\chi)_k^{(1)} \right|\,,
\end{eqnarray}
where we used Eq.~(\ref{QH}).
The above discussion shows that the field-dependent terms in Eq.~(\ref{kappa1})
are neglected relative to the term $\Delta v_\chi/a$, thus ensuring the
validity of the approximation, $\kappa^{(1)} \simeq \Delta v_{\chi}^{(1)}/a$, 
for the modes deep inside the Hubble radius.

The second-order term on the r.h.s. of Eq.~(\ref{kappa2})
is of the order of $H^2 \left| {\delta_k^{(1)}}^2 \right|$ 
by employing the first-order solution (\ref{delves}).
Meanwhile one of the first-order term, $H^2 \delta F_k^{(1)}/F$, 
in Eq.~(\ref{kappa}) has been already estimated in Eq.~(\ref{delFes}).
The former is larger than the latter provided that
\begin{eqnarray}
\label{delcon}
|\delta_k^{(1)}| \gg \frac{Q^2H^2}{M^2/F+k^2/a^2}\,.
\end{eqnarray}
The r.h.s. of Eq.~(\ref{delcon}) is much smaller than unity
because of the condition $\{M^2/F, k^2/a^2 \} \gg H^2$.
One can show that, under the condition (\ref{delcon}), 
other field-dependent first-order terms
on the l.h.s. of Eq.~(\ref{kappa}) can be negligible 
relative to the term $\frac{1}{a^2} \left[ (\nabla v_{\chi}) \cdot
(\nabla v_{\chi})^{,i} \right]_{,i}$.
In summary, the master equation (\ref{basic3}) we have approximately 
derived in the previous section is trustable under 
the conditions (\ref{QH}) and (\ref{delcon}).

\subsection{Second-order perturbations and skewness}

The second-order perturbation $\delta^{(2)}$ satisfies
\begin{eqnarray}
\label{sorder}
\frac{\partial^2 \delta^{(2)}}{\partial t^2}+
2H \frac{\partial \delta^{(2)}}{\partial t}-
4\pi \rho_m  G_{\rm eff}^{(2)} \delta^{(2)}=
4\pi G_{\rm eff}^{(1)} \rho_m \left( \delta^{(1)} \right)^2
+\frac{1}{a^2} \delta_{,i}^{(1)} \Phi_{,i}^{(1)}
+\frac{1}{a^2} \left[ u^{i(1)} u^{j(1)} \right]_{,ij}\,,
\end{eqnarray}
where $G_{\rm eff}^{(2)}\delta^{(2)}=
[\delta^{(2)}-\Delta \delta F^{(2)}/(\rho_m a^2)]/
(8\pi F)$. 
When the growth rate of perturbations is dependent on $k$,
the gravitational constant $G_{\rm eff}^{(2)}$ is generally 
different from $G_{\rm eff}^{(1)}$.
In the following we study the regime of the massless limit ($M \to 0$) 
in which the growth rate of the first-order perturbation is independent of 
$k$, i.e,  $\delta^{(1)}=D(t) \delta_1({\bm x})$ with 
$D(t)=t^{(\sqrt{25+48Q^2}-1)/6}$.
In this regime we have $G_{\rm eff}^{(2)}=G_{\rm eff}^{(1)}
=(1+2Q^2)/8\pi F$, so we simply adopt the notation, $G_{\rm eff}$, 
instead of $G_{\rm eff}^{(1)}$ and $G_{\rm eff}^{(2)}$.
The General Relativistic case ($M \to \infty$) is recovered by 
taking the limit $Q \to 0$.

The first-order solution to ${\bm u}$ can be obtained 
by solving Eq.~(\ref{delves}), i.e., 
$\nabla \cdot {\bm u}^{(1)}=-a\dot{\delta}^{(1)}$. It is given by 
\begin{eqnarray}
\label{vso}
{\bm u}^{(1)}=-\frac{a\dot{D}}{4\pi}
\int \frac{({\bm x}-{\bm x}')\delta_1 ({\bm x}')}
{|{\bm x}-{\bm x}'|^3}\,{\rm d}^3 {\bm x}'
=\frac{a\dot{D}}{4\pi } \Delta_{,i} \,,
\end{eqnarray}
where $\Delta_{,i}$ is a spatial derivative of the quantity: 
\begin{eqnarray}
\Delta ({\bm x}) \equiv
\int \frac{\delta_1 ({\bm x}')}
{|{\bm x}-{\bm x}'|}{\rm d}^3 {\bm x}'\,. 
\end{eqnarray}
This satisfies the relation $\Delta_{,ii}=-4\pi \delta_1 ({\bm x})$.

The last term on the r.h.s. of Eq.~(\ref{sorder}) yields
\begin{eqnarray}
\frac{1}{a^2}
\left[ u^{i(1)} u^{j(1)} \right]_{,ij} &=&
\frac{1}{a^2} \left[u^{i(1)}_{,i} u^{j(1)}_{,j}
+2u^{i(1)}_{,ij} u^{j(1)}
+u^{i(1)}_{,j} u^{j(1)}_{,i} \right] \nonumber \\
&=& \dot{D}^2 
\left[ \delta_1^2
-\frac{1}{2\pi} \delta_{1,j} \Delta_{,j}
+\frac{1}{16\pi^2}\Delta_{,ij} \Delta_{,ij} \right]\,.
\end{eqnarray}
We write the solution of Eq.~(\ref{sorder}) in the form \cite{Marc}
\begin{eqnarray}
\delta^{(2)}=\delta_a^{(2)}+\delta_b^{(2)}\,, 
\end{eqnarray}
where $\delta_a^{(2)}$ and $\delta_b^{(2)}$ satisfy
\begin{eqnarray}
\label{seq1}
& &\frac{\partial^2 \delta_a^{(2)}}{\partial t^2}+
2H \frac{\partial \delta_a^{(2)}}{\partial t}-
4\pi G_{\rm eff} \rho_m \delta_a^{(2)}=
4\pi G_{\rm eff} \rho_m 
D^2 \delta_1^2
+\frac{D}{a^2} \Phi_{,i}^{(1)} \delta_{1,i}\,, \\
\label{seq2}
& & \frac{\partial^2 \delta_b^{(2)}}{\partial t^2}+
2H \frac{\partial \delta_b^{(2)}}{\partial t}-
4\pi G_{\rm eff} \rho_m \delta_b^{(2)}=
\dot{D}^2
\left[ \delta_1^2
-\frac{1}{2\pi} \delta_{1,i} \Delta_{,i}
+\frac{1}{16\pi^2} \Delta_{,ij} \Delta_{,ij} \right]\,.
\end{eqnarray}

Since $\Phi_{,i}^{(1)} =-G_{\rm eff}\rho_m a^2D \Delta_{,i}$ from
Eq.~(\ref{dePhi}), the r.h.s. of Eq.~(\ref{seq1}) is given by 
$4\pi G_{\rm eff}\rho_m D^2
[\delta_1^2-(1/4\pi) \Delta_{,i}\delta_{1,i}]$.
Writing the solution of $\delta_a^{(2)}$ as 
$\delta_a^{(2)}=E_a(t) \delta_a ({\bm x})$, we obtain 
the following equation for the temporal part:
\begin{eqnarray}
\label{Ea}
\ddot{E}_a+2H \dot{E}_a-4\pi G_{\rm eff}\rho_m E_a=
4\pi G_{\rm eff} \rho_m D^2\,,
\end{eqnarray}
where the spatial part is given by $\delta_a ({\bm x})=
\delta_1^2-(1/4\pi) \Delta_{,i}\delta_{1,i}$.
Expressing the solution of Eq.~(\ref{seq2})
in the form $\delta_b^{(2)}=E_b(t) \delta_b ({\bm x})$, 
we get 
\begin{eqnarray}
\label{Eb}
\ddot{E}_b+2H \dot{E}_b
-4\pi G_{\rm eff}\rho_m E_b=\dot{D}^2\,,
\end{eqnarray}
and $\delta_b ({\bm x})=
\delta_1^2-(1/2\pi) \Delta_{,i}\delta_{1,i}
+(1/16\pi^2) \Delta_{,ij}\Delta_{,ij}$.

We then find the following solution 
for second-order perturbations:
\begin{eqnarray}
\delta^{(2)}(t, {\bm x}) &=&
E_a(t) \left[ \delta_1^2- \frac{1}{4\pi}
\Delta_{,i}\delta_{1,i} \right]+
E_b(t) \left[ 
\delta_1^2-\frac{1}{2\pi} \Delta_{,i}\delta_{1,i}
+\frac{1}{16\pi^2} \Delta_{,ij}\Delta_{,ij} \right]\,, \nonumber \\
&=& \frac{D^2+E_a}{2} \delta_1^2-\frac{D^2}{4\pi}
\Delta_{,i} \delta_{1,i}+\frac{D^2-E_a}{32\pi^2}
\Delta_{,ij} \Delta_{,ij}\,.
\end{eqnarray}
In the second line we employed the fact that $E_a$ and $E_b$ 
are related each other via the relation $E_a+2E_b=D^2$.

We assume that  the initial distribution of perturbations is Gaussian
so that it is described by an auto-correlation function $\xi ({\bm x})$
satisfying
\begin{eqnarray}
& &
\langle \delta ({\bm x}) \rangle=0\,,\quad
\langle \delta ({\bm x}_1) \delta ({\bm x}_2) \rangle
=\xi (|{\bm x}_1-{\bm x}_2|)\,,\quad
\langle \delta ({\bm x}_1) \delta ({\bm x}_2) 
\delta ({\bm x}_3) \rangle=0\,, \nonumber \\
& &
\langle \delta ({\bm x}_1) \delta ({\bm x}_2) 
\delta ({\bm x}_3) \delta ({\bm x}_4) \rangle
=\xi (|{\bm x}_1-{\bm x}_2|) \xi (|{\bm x}_3-{\bm x}_4|)
+\xi (|{\bm x}_1-{\bm x}_3|) \xi (|{\bm x}_2-{\bm x}_4|)
+\xi (|{\bm x}_1-{\bm x}_4|) \xi (|{\bm x}_2-{\bm x}_3|)\,.
\end{eqnarray}
Since $\langle (\delta^{(1)})^3 \rangle=0$, the quantity 
$\langle \delta^3 \rangle$ is given by 
$\langle \delta^3 \rangle=3\langle (\delta^{(1)})^2
\delta^{(2)} \rangle$ to the lowest order.
We then have
\begin{eqnarray}
\label{del3}
\langle \delta^3 \rangle=
\frac32 D^2(D^2+E_a) \langle \delta_1^4 \rangle
-\frac{3}{4\pi}D^4 \langle \delta_1^2 \delta_{1,i} \Delta_{,i}
\rangle+\frac{3}{32\pi^2} D^2(D^2-E_a)
\langle \delta_1^2 \Delta_{,ij}\Delta_{,ij} \rangle\,.
\end{eqnarray}

Since the each ensemble average in Eq.~(\ref{del3}) satisfies 
the relations 
$ \langle \delta_1^4 \rangle=3\xi(0)^2$,
$\langle \delta_1^2 \delta_{1,i} \Delta_{,i}
\rangle=4\pi \xi(0)^2$ and 
$\langle \delta_1^2 \Delta_{,ij}\Delta_{,ij} \rangle
=\frac{80\pi^2}{3} \xi(0)^2$ \cite{Peebles},
we obtain the skewness
\begin{eqnarray}
S_3 \equiv \frac{\langle \delta^3 \rangle}
{\langle \delta^2 \rangle^2}=4+2\frac{E_a}{D^2}\,,
\end{eqnarray}
where we used $ \langle \delta_1^2 \rangle=\xi(0)^2$.
Hence the skewness is determined by the second-order 
growth rate $E_a$ relative to the squared of 
the first-order growth rate $D$.

Equation (\ref{Ea}) can be written as
\begin{eqnarray}
\label{Easi}
E_a''+\left(\frac12-\frac32 w_{\rm eff} \right)E_a'-
12\pi F G_{\rm eff} \Omega_m E_a=
12\pi F G_{\rm eff} \Omega_m D^2\,.
\end{eqnarray}
Recall that $G_{\rm eff}=(1+2Q^2)/8\pi F$
in the limit $M \to 0$.
During the matter era realized by the point (M2)
we have $\Omega_m \simeq 0$ and $w_{\rm eff} \simeq 0$, 
in which case Eq.~(\ref{Easi}) reduces \footnote{Note that the
skewness was calculated in Ref.~\cite{BD2} when the matter era is 
realized by the point (M1). As we already mentioned, the point (M1)
can not be used for the matter era when the coupling $|Q|$ is 
of the order of unity.}
\begin{eqnarray}
\label{Easi2}
E_a''+\frac12 E_a'-\frac32 (1+2Q^2)E_a=
\frac32 (1+2Q^2) D^2\,.
\end{eqnarray}
Using the first-order solution $D=e^{\frac14 (\sqrt{25+48Q^2}-1)N}$,
we get the following special solution for Eq.~(\ref{Easi2}):
\begin{eqnarray}
E_a=\frac{6(1+2Q^2)}{19+36Q^2-\sqrt{25+48Q^2}}
e^{\frac12 (\sqrt{25+48Q^2}-1)N}\,.
\end{eqnarray}
Hence the skewness in the regime of the massless limit 
is given by 
\begin{eqnarray}
\label{anaes}
S_3=\frac{4[22+42Q^2-\sqrt{25+48Q^2}]}
{19+36Q^2-\sqrt{25+48Q^2}}\,.
\end{eqnarray}
The General Relativistic case is recovered by taking the limit 
$Q \to 0$:
\begin{eqnarray}
S_3=34/7\,.
\end{eqnarray}
This agrees with the skewness in the Einstein-de Sitter 
Universe \cite{Peebles} 
(pressureless matter without cosmological constant).

%%%%%%%%%%%%%%%%%%%%%%%%%%%%
\begin{figure}
\includegraphics[height=3.2in,width=3.2in]{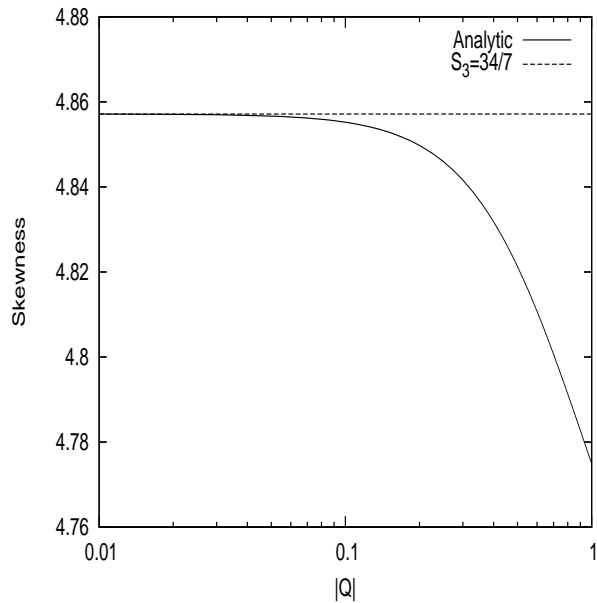}
\caption{
The analytic estimation (\ref{anaes}) of the skewness 
during the matter-dominated epoch in the regime of 
the massless limit ($M \to 0$).
With the increase of $|Q|$ the skewness gets smaller 
compared to the value $S_3=34/7$ 
in the Einstein-de Sitter Universe.
However the difference from the Einstein-de Sitter case
is only less than 1.7 \% for $|Q| \le 1$.
}
\label{ana}
\end{figure}
%%%%%%%%%%%%%%%%%%%%%%%%%%%%

In Fig.~\ref{ana} we plot the analytic value (\ref{anaes})
as a function of $|Q|$.
The skewness shows some difference compared to the 
Einstein-de Sitter value $34/7$ for $|Q|>0.1$.
When $|Q|=1$ we have $S_3=4.775$, which is different from 
the value $34/7$ only by 1.7 \%.
For the potential (\ref{potential}) the first-order perturbation
$\delta_k^{(1)}$ evolves from the regime $M^2/F \gg k^2/a^2$
to the regime  $M^2/F \ll k^2/a^2$ for the modes 
relevant to large-scale structure.
Hence the skewness tends to evolves from the value $34/7$ to 
the asymptotic value given in Eq.~(\ref{anaes}).
The estimation (\ref{anaes}) has been derived by neglecting 
the transient phase around the redshift $z_k$.
Since this transition occurs quickly for the models that 
satisfy local gravity constraints 
($p>0.7$) \cite{Star07,Tsuji08,TUMTY},
it is unlikely that the skewness is altered significantly
by the presence of this transient phase.

The estimation (\ref{anaes}) does not take into account the evolution 
in the late-time accelerated epoch.
In the $\Lambda$CDM model the numerical analysis shows that the 
skewness increases a bit during the accelerated phase from the 
value $34/7~(=4.857)$ to the present value $4.865$ (at $\Omega_m=0.28$).
This corresponds to the growth of 0.16 \% only.
We have checked that this situation does not change much even in 
the presence of the coupling $Q$.
Hence the difference of the present values of the skewness from that 
in the $\Lambda$CDM model is only less than a few percent.
This shows that the skewness provides a robust 
prediction for the picture of gravitational instability 
from Gaussian initial conditions, 
including scalar-tensor models 
with large couplings ($|Q| \lesssim 1$).

%%%%%%%%%%%%%%%%%%%%%%%%%%
\section{Conclusions}
\label{conclude}
%%%%%%%%%%%%%%%%%%%%%%%%%%

In this paper we have studied the evolution of second-order
matter density perturbations in a class of modified gravity models that 
satisfy local gravity constraints.
We have considered the scalar-tensor action 
(\ref{action}), which is equivalent to Brans-Dicke
action (\ref{action0}) with the correspondence
$1/(2Q^2)=3+2\omega_{\rm BD}$.
In the presence of a field potential it is possible to satisfy  
local gravity constraints (LGC) even when $|Q|$
is of the order of unity.
In fact the potential (\ref{potential}) is designed to have a 
large mass in the region of high density for the consistency 
with LGC. This covers the models proposed by 
Hu and Sawicki \cite{Hu07} and Starobinsky \cite{Star07} 
in the context of $f(R)$ gravity ($Q=-1/\sqrt{6}$).

Starting from second-order relativistic equations 
of cosmological perturbations, we have derived
the equation (\ref{basic3}) of matter density 
fluctuations approximately. 
In so doing we employed the approximation that 
first-order perturbations in the scalar field $\phi$
is neglected relative to second-order matter and velocity 
perturbations. This is valid under the conditions
(\ref{QH}) and (\ref{delcon}), both of which
can be naturally satisfied 
for the values of $Q$ we are interested in 
($0.1 \lesssim |Q| \lesssim 1$).
Compared to the $\Lambda$CDM model, the effective 
gravitational constant $G_{\rm eff}$ is subject to change
at the late epoch of the matter era.
This leads to the larger growth rate of first-order 
matter perturbations ($\delta_k^{(1)} \propto t^{(\sqrt{25+48Q^2}-1)/6}$)
compared to the standard case ($\delta_k^{(1)} \propto t^{2/3}$).

The skewness of matter distributions is determined by 
the second-order growth factor $E_a$ relative to 
the squared of the first-order growth factor $D$.
In the ``scalar-tensor regime'' where the effective gravitational 
constant is given by $G_{\rm eff} \simeq (1+2Q^2)/8\pi F$, 
we have derived the analytic expression (\ref{anaes}) of 
the skewness in the matter-dominated epoch.
In the ``General Relativistic regime'' where
$G_{\rm eff} \simeq 1/8\pi F \simeq G$,
we have reproduced the standard value $S_3=34/7$ in the 
Einstein-de Sitter Universe.
In modified gravity models with $|Q| \lesssim 1$, the 
analytic value (\ref{anaes}) of the skewness in the asymptotic regime 
of the matter era is different from the value $34/7$
only less than a few percent.
Even if we take into account the evolution of perturbations
during the accelerated phase, the difference of the skewness
relative to the $\Lambda$CDM model remains to be small.
The above result comes from the fact that the ratio of the second-order
growth rate relative to the first-order one has a weak 
dependence on the coupling $Q$.

When $|Q|={\cal O} (1)$ the growth rate of first-order matter perturbations
is significantly different from that in the $\Lambda$CDM model.
This gives rise to large modifications to the matter power spectrum 
as well as to the convergence spectrum in weak lensing, while 
the skewness is hardly distinguishable from that in $\Lambda$CDM model.
This fact can be useful to discriminate large coupling 
scalar-tensor models among many other dark energy 
models from future high-precision observations.

%%%%%%%%%%%%%%%%%%%%%
\section*{ACKNOWLEDGEMENTS}
This work was partially supported by JSPS
Grant-in-Aid for Scientific Research
No.~30318802 (ST).
%%%%%%%%%%%%%%%%%%%%%

%%%%%%%%%%%%%%%%%%%%%
\section*{Appendix}
%%%%%%%%%%%%%%%%%%%%%

In this Appendix, we estimate the order of second-order terms that we have 
neglected in Eqs.~(\ref{delphi2}) and (\ref{delF2}).
The dominant contribution of such second-order terms comes from 
the third-derivative of the potential, i.e., $V_{,\phi \phi \phi} \delta \phi^2$ \cite{HN}.
Compared to the field-mass dependent term $V_{,\phi \phi} \delta \phi$
on the l.h.s. of Eq.~(\ref{delphi2}), we have 
$(V_{,\phi \phi \phi} \delta \phi^2)/(V_{,\phi \phi} \delta \phi) \approx
-\delta \phi/\phi$ for the potential (\ref{potential}) under the condition $|Q \phi| \ll 1$.

Since the field stays around the instantaneous minimum given in Eq.~(\ref{phim}),
the field $\phi$ can be estimated as 
\begin{eqnarray}
\phi \simeq 3(1-p)Q \frac{H^2}{M^2}\,,
\end{eqnarray}
where we used Eq.~(\ref{mass}).
In deriving this, we have also employed the approximate relation $\rho_m \approx 3H^2$. 
Note that the order of $\rho_m$ is not different from $3H^2$ even at the present epoch.
Meanwhile, from Eq.~(\ref{delF3}), the first-order perturbation $\delta \phi_k^{(1)}$
in the Fourier space during the matter era is given by 
\begin{eqnarray}
\delta \phi_k^{(1)} \simeq -\frac{3Q H^2}{M^2+k^2/a^2} \delta_k^{(1)}\,.
\end{eqnarray}
Hence we obtain the ratio
\begin{eqnarray}
\frac{\delta \phi_k^{(1)}}{\phi} \simeq 
-\frac{1}{1-p} \frac{M^2}{M^2+k^2/a^2}
 \delta_k^{(1)}\,,
\end{eqnarray}
which shows that $|\delta \phi_k^{(1)}/\phi| \ll 1$ for $\delta_k^{(1)} \ll 1$.

The presence of the second-order term $V_{,\phi \phi \phi} \delta \phi^2$
gives rise to a correction of the order $M^2 \delta \phi_k^{(1)}/\phi$ 
to the mass squared $M^2$ in Eq.~(\ref{Geff2}).
In two asymptotic regimes (i) $M^2 \gg k^2/a^2$ and 
(ii) $M^2 \ll k^2/a^2$, this appears only as next-order 
corrections to the small expansion parameters $(k^2/a^2)/M^2$ [regime (i)] 
and $M^2/(k^2/a^2)$ [regime (ii)].

In the regime $M^2 \gg k^2/a^2$ the correction from the term  
$V_{,\phi \phi \phi} \delta \phi^2$ to the effective 
gravitational constant $G_{\rm eff}^{(1)}$ is estimated as 
$\delta G_{\rm eff}^{(1)} \approx Q^2 (k^2/a^2 M^2) \delta \phi_k^{(1)}/\phi$.
This gives the correction to the third term on the l.h.s. of 
Eq.~(\ref{kappa3}) in the Fourier space:
\begin{eqnarray}
\left| 4 \pi \rho_m\,\delta G_{\rm eff}^{(1)}\,\delta_k^{(1)} \right|
\approx \frac{Q^2}{1-p} \frac{k^2/a^2}{M^2} H^2 {\delta_k^{(1)}}^2\,,
\end{eqnarray}
which is much smaller than the second-order term on the r.h.s. of 
Eq.~(\ref{kappa3}) that is of the order of $H^2 {\delta_k^{(1)}}^2$.

In the regime $M^2 \ll k^2/a^2$ we have 
$\delta G_{\rm eff}^{(1)} \approx Q^2 M^2(a^2/k^2)
\delta \phi_k^{(1)}/\phi$ and hence
\begin{eqnarray}
\left| 4 \pi \rho_m\,\delta G_{\rm eff}^{(1)}\,\delta_k^{(1)} \right|
\approx \frac{Q^2}{1-p} \left( \frac{M^2}{k^2/a^2} \right)^2 
H^2 {\delta_k^{(1)}}^2\,,
\end{eqnarray}
which is again much smaller than the r.h.s. of Eq.~(\ref{kappa3}).

The above estimation shows that neglecting second-order terms 
in Eqs.~(\ref{delphi2}) and (\ref{delF2}) is justified.

%%%%%%%%%%%%%%%%%


\begin{thebibliography}{10}
%%%%%%%%%%%%%%%%%

\bibitem{Perl}
S.~Perlmutter {\it et al.},
%``Measurements of Omega and Lambda
%from 42 High-Redshift Supernovae,''
Astrophys.\ J.\  {\bf 517}, 565 (1999);
A.~G.~Riess {\it et al.},
%``Observational Evidence from Supernovae for
%an Accelerating Universe and a Cosmological Constant,''
Astron.\ J.\  {\bf 116}, 1009 (1998);
Astron.\ J.\  {\bf 117}, 707 (1999);
M.~Tegmark \textit{et al.}, 
%``Cosmological parameters from SDSS and WMAP,''
Phys.\ Rev.\ D \textbf{69}, 103501 (2004); 
U.~Seljak \textit{et al.}, Phys.\ Rev.\ D \textbf{71}, 103515 (2005);
D.~N.~Spergel \textit{et al.}, 
Astrophys.\ J.\ Suppl.\ \textbf{148}, 175 (2003); 
E.~Komatsu {\it et al.} 
arXiv:0803.0547 [astro-ph].

\bibitem{review}
V.~Sahni and A.~A.~Starobinsky, Int.\ J.\ Mod.\ Phys.\ D \textbf{9},
373 (2000); V.~Sahni, Lect.\ Notes Phys.\ {} \textbf{653}, 141 (2004);
S.~M.~Carroll, Living Rev.\ Rel.\ {} \textbf{4}, 1 (2001);
T.~Padmanabhan, Phys.\ Rept.\ {} \textbf{380}, 235 (2003);
P.~J.~E.~Peebles and B.~Ratra, Rev.\ Mod.\
Phys.\ {} \textbf{75}, 559 (2003);
S.~Nojiri and S.~D.~Odintsov,
Int.\ J.\ Geom.\ Meth.\ Mod.\ Phys.\  {\bf 4}, 115 (2007).

\bibitem{CST}
E.~J.~Copeland, M.~Sami and S.~Tsujikawa,
%``Dynamics of dark energy,''
Int.\ J.\ Mod.\ Phys.\  D {\bf 15}, 1753 (2006).

\bibitem{fR}
S.~Capozziello, Int. J. Mod. Phys. {\bf D 11}, 483, (2002);
S.~Capozziello, V.~F.~Cardone, S.~Carloni and A.~Troisi,
Int. J. Mod. Phys. {\bf D}, 12, 1969 (2003);
S.~M.~Carroll, V.~Duvvuri, M.~Trodden and M.~S.~Turner,
%``Is cosmic speed-up due to new gravitational physics?,''
Phys.\ Rev.\  D {\bf 70}, 043528 (2004);
S.~Nojiri and S.~D.~Odintsov,
Phys. Rev. {\bf D 68}, 12352, (2003).

\bibitem{stensor}
L.~Amendola,
%``Scaling solutions in general non-minimal coupling theories,''
Phys.\ Rev.\  D {\bf 60}, 043501 (1999);
J.~P.~Uzan,
%``Cosmological scaling solutions of non-minimally
%coupled scalar fields,''
Phys.\ Rev.\  D {\bf 59}, 123510 (1999);
T.~Chiba,
%``Quintessence, the gravitational constant, and gravity,''
Phys.\ Rev.\ D {\bf 60}, 083508 (1999);
N.~Bartolo and M.~Pietroni,
%``Scalar tensor gravity and quintessence,''
Phys.\ Rev.\ D {\bf 61} 023518 (2000);
F.~Perrotta, C.~Baccigalupi and S.~Matarrese,
%``Extended quintessence,''
Phys.\ Rev.\ D {\bf 61}, 023507 (2000);
C.~Baccigalupi, S.~Matarrese and F.~Perrotta,
%``Tracking extended quintessence,''
Phys.\ Rev.\ D {\bf 62}, 123510 (2000).

\bibitem{Boi}
B.~Boisseau, G.~Esposito-Farese, D.~Polarski and A.~A.~Starobinsky,
Phys.\ Rev.\ Lett.\  {\bf 85}, 2236 (2000);
G.~Esposito-Farese and D.~Polarski,
%``Scalar-tensor gravity in an accelerating universe,''
Phys.\ Rev.\  D {\bf 63}, 063504 (2001).

\bibitem{brane}
G.~R.~Dvali, G.~Gabadadze and M.~Porrati,
%``4D gravity on a brane in 5D Minkowski space,''
Phys.\ Lett.\  B {\bf 485}, 208 (2000);
V.~Sahni and Y.~Shtanov,
%``Braneworld models of dark energy,''
JCAP {\bf 0311}, 014 (2003).

\bibitem{Star07}
A.~A.~Starobinsky,
%``Disappearing cosmological constant in f(R) gravity,''
JETP Lett.\  {\bf 86}, 157 (2007).

\bibitem{staper}
Y.~S.~Song, W.~Hu and I.~Sawicki,
%``The large scale structure of f(R) gravity,''
Phys.\ Rev.\  D {\bf 75}, 044004 (2007);
I.~Sawicki and W.~Hu,
%``Stability of Cosmological Solution in f(R) Models of Gravity,''
Phys.\ Rev.\  D {\bf 75}, 127502 (2007);
S.~M.~Carroll, I.~Sawicki, A.~Silvestri and M.~Trodden,
%``Modified-Source Gravity and Cosmological Structure Formation,''
New J.\ Phys.\  \textbf{8}, 323 (2006);
R.~Bean, D.~Bernat, L.~Pogosian, A.~Silvestri and M.~Trodden,
%``Dynamics of Linear Perturbations in f(R) Gravity,''
Phys.\ Rev.\  D \textbf{75}, 064020 (2007);
Y.~S.~Song, H.~Peiris and W.~Hu,
%``Cosmological Constraints on f(R) Acceleration Models,''
Phys.\ Rev.\  D {\bf 76}, 063517 (2007);
L.~Pogosian and A.~Silvestri,
%``The pattern of growth in viable f(R) cosmologies,''
Phys.\ Rev.\  D {\bf 77}, 023503 (2008).

\bibitem{AGPT}
L.~Amendola, R.~Gannouji, D.~Polarski and S.~Tsujikawa,
%``Conditions for the cosmological viability of f(R) dark energy models,''
Phys.\ Rev.\  D {\bf 75}, 083504 (2007).

\bibitem{Nava}
I.~Navarro and K.~Van Acoleyen,
%``f(R) actions, cosmic acceleration and local tests of gravity,''
JCAP {\bf 0702}, 022 (2007);
T.~Faulkner, M.~Tegmark, E.~F.~Bunn and Y.~Mao,
%``Constraining f(R) gravity as a scalar tensor theory,''
Phys.\ Rev.\  D {\bf 76}, 063505 (2007).

\bibitem{CT}
S.~Capozziello and S.~Tsujikawa,
%``Solar system and equivalence principle constraints 
%on $f(R)$ gravity by chameleon approach,''
Phys.\ Rev.\  D {\bf 77}, 107501 (2008).

\bibitem{TUT}
S.~Tsujikawa, K.~Uddin and R.~Tavakol,
%``Density perturbations in f(R) gravity theories in metric and
%Palatini formalisms,''
Phys.\ Rev.\  D {\bf 77}, 043007 (2008).

\bibitem{Brax}
P.~Brax, C.~van de Bruck, A.~C.~Davis and D.~J.~Shaw,
%``f(R) Gravity and Chameleon Theories,''
arXiv:0806.3415 [astro-ph].

\bibitem{Faraoni}
V.~Faraoni,
%``de Sitter attractors in generalized gravity,''
Phys.\ Rev.\  D {\bf 70}, 044037 (2004);
Phys.\ Rev.\  D {\bf 72}, 061501 (2005).

\bibitem{Hu07}
W.~Hu and I.~Sawicki,
%``Models of f(R) Cosmic Acceleration that Evade Solar-System Tests,''
Phys.\ Rev.\  D {\bf 76}, 064004 (2007).

\bibitem{Li}
B.~Li and J.~D.~Barrow,
%``The Cosmology of f(R) Gravity in Metric Variational Approach,''
Phys.\ Rev.\  D {\bf 75}, 084010 (2007).

\bibitem{AT}
L.~Amendola and S.~Tsujikawa,
%``Phantom crossing, equation-of-state singularities, and local gravity
%constraints in $f(R)$ models,''
Phys.\ Lett.\  B {\bf 660}, 125 (2008).

\bibitem{Appleby}
S.~A.~Appleby and R.~A.~Battye,
%``Do consistent $F(R)$ models mimic General Relativity plus $\Lambda$?,''
Phys.\ Lett.\  B {\bf 654}, 7 (2007));
arXiv:0803.1081 [astro-ph].

\bibitem{Tsuji08}
S.~Tsujikawa,
%``Observational signatures of f(R) dark energy models that satisfy
%cosmological and local gravity constraints,''
Phys.\ Rev.\  D {\bf 77}, 023507 (2008).

\bibitem{NO07}
S.~Nojiri and S.~D.~Odintsov,
Phys.\ Lett.\  B {\bf 657}, 238 (2007);
G.~Cognola {\it et al.},
Phys.\ Rev.\  D {\bf 77}, 046009 (2008).

\bibitem{APT}
L.~Amendola, D.~Polarski and S.~Tsujikawa,
%``Are f(R) dark energy models cosmologically viable ?,''
Phys.\ Rev.\ Lett.\  {\bf 98}, 131302 (2007);
Int.\ J.\ Mod.\ Phys.\  D {\bf 16}, 1555 (2007).

\bibitem{KW}
J.~Khoury and A.~Weltman,
%``Chameleon fields: Awaiting surprises for tests of gravity in space,''
Phys.\ Rev.\ Lett.\  {\bf 93}, 171104 (2004);
J.~Khoury and A.~Weltman,
%``Chameleon cosmology,''
Phys.\ Rev.\  D {\bf 69}, 044026 (2004);
P.~Brax, C.~van de Bruck, A.~C.~Davis, J.~Khoury and A.~Weltman,
%``Detecting dark energy in orbit: The cosmological chameleon,''
Phys.\ Rev.\  D {\bf 70}, 123518 (2004);
D.~F.~Mota and D.~J.~Shaw,
%``Strongly coupled chameleon fields: New horizons in scalar field theory,''
Phys.\ Rev.\ Lett.\  {\bf 97}, 151102 (2006);
Phys.\ Rev.\  D {\bf 75}, 063501 (2007)s.

\bibitem{coupled}
L.~Amendola,
%``Coupled quintessence,''
Phys.\ Rev.\  D {\bf 62}, 043511 (2000).

\bibitem{TT}
S.~Tsujikawa and T.~Tatekawa,
%``The effect of modified gravity on weak lensing,''
Phys.\ Lett.\  B {\bf 665}, 325 (2008).

\bibitem{Schmidt}
F.~Schmidt,
%``Weak Lensing Probes of Modified Gravity,''
Phys.\ Rev.\  D {\bf 78}, 043002 (2008).

\bibitem{TUMTY}
S.~Tsujikawa, K.~Uddin, S.~Mizuno, R.~Tavakol and J.~Yokoyama,
%``Constraints on scalar-tensor models of dark energy from
% observational and local gravity tests,''
Phys.\ Rev.\  D {\bf 77}, 103009 (2008).

\bibitem{BD}
C.~Brans and R.~H.~Dicke,
%``Mach's principle and a relativistic theory of gravitation,''
Phys.\ Rev.\  {\bf 124}, 925 (1961).

% Skewness papers

\bibitem{Ber}
F.~Bernardeau, S.~Colombi, E.~Gaztanaga and R.~Scoccimarro,
%``Large-scale structure of the universe and 
%cosmological perturbation theory,''
Phys.\ Rept.\  {\bf 367}, 1 (2002).

\bibitem{Peebles}
P.~J.~E.~Peebles, ``Large-Scale Structure of the Universe'',
Princeton University Press (1980).

\bibitem{Marc}
M.~Kamionkowski and A.~Buchalter,
%``Weakly Nonlinear Clustering for Arbitrary Expansion Histories,''
Astrophys.\ J.\  {\bf 514}, 7 (1999).

\bibitem{Benabed}
K.~Benabed and F.~Bernardeau,
%``Testing quintessence models with large-scale structure growth,''
Phys.\ Rev.\  D {\bf 64}, 083501 (2001).

\bibitem{Kofman}
L.~Kofman, E.~Bertschinger, J.~M.~Gelb, A.~Nusser and A.~Dekel,
%``Evolution of one point distributions from Gaussian initial fluctuations,''
Astrophys.\ J.\  {\bf 420}, 44 (1994).

\bibitem{curved}
F.~R.~Bouchet, R.~Juszkiewicz, S.~Colombi and R.~Pellat,
%``Weakly nonlinear gravitational instability for arbitrary Omega,''
Astrophys.\ J.\  {\bf 394}, L5 (1992).

\bibitem{Mul}
T.~Multamaki, E.~Gaztanaga and M.~Manera,
%``Large scale structure in non-standard cosmologies,''
Mon.\ Not.\ Roy.\ Astron.\ Soc.\  {\bf 344}, 761 (2003).

\bibitem{Freese}
K.~Freese and M.~Lewis,
%``Cardassian Expansion: a Model in which the Universe is Flat, Matter
%Dominated, and Accelerating,''
Phys.\ Lett.\  B {\bf 540}, 1 (2002).

\bibitem{Lue}
A.~Lue, R.~Scoccimarro and G.~Starkman,
%``Differentiating between Modified Gravity and Dark Energy,''
Phys.\ Rev.\  D {\bf 69}, 044005 (2004).

\bibitem{Luca04}
L.~Amendola and C.~Quercellini,
%``Skewness as a test of the equivalence principle,''
Phys.\ Rev.\ Lett.\  {\bf 92}, 181102 (2004).

\bibitem{BD2}
E.~Gaztanaga and J.~A.~Lobo,
%``Non-Linear gravitational growth of large scale structures inside and
%outside standard Cosmology,''
Astrophys.\ J.\  {\bf 548}, 47 (2001).

\bibitem{Hoyle}
C.~D.~Hoyle {\it et al.,}
%``Sub-millimeter tests of the gravitational inverse-square law,''
Phys.\ Rev.\ D \textbf{70}, 042004 (2004).

\bibitem{Chiba}
T.~Chiba,
%``1/R gravity and scalar-tensor gravity,''
Phys.\ Lett.\  B {\bf 575}, 1 (2003).

%%%%%

\bibitem{metper}
J.~M.~Bardeen,
%``Gauge Invariant Cosmological Perturbations,''
Phys.\ Rev.\  D {\bf 22}, 1882 (1980);
H.~Kodama and M.~Sasaki,
%``Cosmological Perturbation Theory,''
Prog.\ Theor.\ Phys.\ Suppl.\  {\bf 78}, 1 (1984);
V.~F.~Mukhanov, H.~A.~Feldman and R.~H.~Brandenberger,
Phys.\ Rept.\  {\bf 215}, 203 (1992);
B.~A.~Bassett, S.~Tsujikawa and D.~Wands,
%``Inflation dynamics and reheating,''
Rev.\ Mod.\ Phys.\  {\bf 78}, 537 (2006).

\bibitem{Noh}
H.~Noh and J.~c.~Hwang,
%``Second-order perturbations of the Friedmann world model,''
Phys.\ Rev.\  D {\bf 69}, 104011 (2004).

\bibitem{HN}
J.~c.~Hwang and H.~Noh,
Phys.\ Rev.\  D {\bf 65}, 023512 (2002);
Phys.\ Rev.\  D {\bf 71}, 063536 (2005).

\bibitem{Tsuji07}
S.~Tsujikawa,
%``Matter density perturbations and effective gravitational constant in
%modified gravity models of dark energy,''
Phys.\ Rev.\  D {\bf 76}, 023514 (2007).

\end{thebibliography}
\end{document}